\documentclass[lettersize,journal]{IEEEtran}
\usepackage{amsmath,amsfonts}
\usepackage[numbers,compress]{natbib}
\usepackage{tabularx}  
\usepackage{booktabs}  
\usepackage{pifont}
\usepackage{algorithm}
\usepackage[noend]{algpseudocode}
\usepackage{array}
\usepackage[labelfont=scriptsize,textfont=scriptsize]{subfig}
\usepackage{textcomp}
\usepackage{soul,color,xcolor}
\usepackage{enumitem}
\usepackage{stfloats}
\usepackage{tcolorbox}
\usepackage{url}
\usepackage{multirow}
\usepackage{threeparttable}
\usepackage{booktabs}
\usepackage{multirow}
\usepackage{verbatim}
\usepackage{graphicx}
\usepackage{xspace}
\usepackage{url}
\usepackage{bbm, dsfont}
\usepackage{csquotes}
\usepackage{acro}
\usepackage{bbding}

\newcommand{\sys}{\textsf{\textsc{DynaShard}}\xspace}
\makeatletter
\def\hlinewd#1{%
	\noalign{\ifnum0=`}\fi\hrule \@height #1 \futurelet
	\reserved@a\@xhline}
\makeatother

\usepackage{hyperref}
\hypersetup{
	colorlinks=true,
	linkcolor=blue,
	filecolor=blue,      
	urlcolor=blue,
	citecolor=blue,
	pdfpagemode=FullScreen,
}

\renewcommand{\paragraph}[1]{%
\textbf{#1.}
}

\DeclareAcronym{bft}{
    short = BFT,
    long = byzantine fault tolerance
}

\DeclareAcronym{tps}{
    short = TPS,
    long = transactions per second
}

\DeclareAcronym{pbft}{
    short = PBFT,
    long = practical byzantine fault tolerance
}

\DeclareAcronym{snark}{
    short = SNARK,
    long = succinct non-interactive arguments of knowledge
}

\DeclareAcronym{ftsbs}{
    short = FTSBS,
    long = fast transaction scheduling in blockchain sharding
}

\DeclareAcronym{iot}{
    short = IoT,
    long = Internet of Things
}

\DeclareAcronym{zkp}{
    short = ZKP,
    long = Zero-knowledge proof
}

\DeclareAcronym{rpcs}{
    short = RPCs,
    long = remote procedure calls
}

\DeclareAcronym{pow}{
    short = PoW,
    long = proof of work
}

\begin{document}

\title{\sys: Secure and Adaptive Blockchain Sharding Protocol with Hybrid Consensus and Dynamic Shard Management
    \author{Ao Liu$^1$, Jing Chen$^1$, Kun He$^1$, Ruiying Du$^1$\Envelope, Jiahua Xu$^2$, Cong Wu$^3$, Yebo Feng$^3$, Teng Li$^4$, and Jianfeng Ma$^4$\\
    	$^1$\emph{Wuhan University, China}, 	$^2$\emph{University College London, and DLT Science Foundation, UK}\\
    	$^3$\emph{Nanyang Technological University, Singapore}, $^4$\emph{Xidian University, China}\\
  			\texttt{\small \{liuao6,chenjing,hekun,duraying\}@whu.edu.cn, jiahua.xu@ucl.ac.uk, \{cong.wu,yebo.feng\}@ntu.edu.sg, litengxidian@gmail.com, jfma@mail.xidian.edu.cn}}}

\maketitle
\begin{abstract}
    Blockchain sharding has emerged as a promising solution to the scalability challenges in traditional blockchain systems by partitioning the network into smaller, manageable subsets called shards.
    Despite its potential, existing sharding solutions face significant limitations in handling dynamic workloads, ensuring secure cross-shard transactions, and maintaining system integrity.
    To address these gaps, we propose \sys, a dynamic and secure cross-shard transaction processing mechanism designed to enhance blockchain sharding efficiency and security.
    \sys combines adaptive shard management, a hybrid consensus approach, plus an efficient state synchronization and dispute resolution protocol.
    Our performance evaluation, conducted using a robust experimental setup with real-world network conditions and transaction workloads, demonstrates \sys's superior throughput, reduced latency, and improved shard utilization compared to the \ac{ftsbs} method.
    Specifically, \sys achieves up to a 42.6\% reduction in latency and a 78.77\% improvement in shard utilization under high transaction volumes and varying cross-shard transaction ratios.
    These results highlight \sys's ability to outperform state-of-the-art sharding methods, ensuring scalable and resilient blockchain systems.
    We believe that \sys's innovative approach will significantly impact future developments in blockchain technology, paving the way for more efficient and secure distributed systems.
\end{abstract}

\begin{IEEEkeywords}
    Blockchain sharding, cross-shard transactions, secure consensus, state synchronization, scalability
\end{IEEEkeywords}

\section{Introduction}
Blockchain technology has emerged as a groundbreaking innovation that has transformed the landscape of digital transactions, offering a decentralized, transparent, and secure framework for various applications~\cite{wu2024semantic,liang2024vulseye,liang2024towards,liang2024ponziguard}, including \ac{iot}-based systems \cite{reyna2018blockchain,lin2024fedsn,fang2024ic3m,lin2024efficient}. The core principles of blockchain, such as immutability, transparency, and distributed consensus, have the potential to revolutionize \ac{iot} networks by enhancing security, data integrity, and trust among devices and participants \cite{yadav2023evolution}. However, the scalability of blockchain systems remains a significant challenge, particularly for \ac{iot} applications that require high transaction throughput and low latency \cite{rao2024scalability,lin2024split,fang2024automated,lin2023pushing}. The increasing number of \ac{iot} devices and transactions leads to longer confirmation time, higher transaction fees, and reduced overall system performance \cite{alharby2023transaction}. To address this scalability challenge, researchers have proposed various solutions, including off-chain scaling techniques like payment channels \cite{guo2023cross, jia2023cross, qin2023blindhub} and sidechains \cite{li2023review, tran2023enhancing, gai2023secure}, as well as on-chain scaling approaches like sharding \cite{wang2019sok}. By improving scalability, these solutions can enable the seamless integration of blockchain technology with \ac{iot} systems, ensuring efficient and secure management of \ac{iot} data and transactions.

\paragraph{Existing efforts}
Blockchain sharding has gained significant attention as a promising on-chain scaling solution that aims to improve the throughput and latency of blockchain systems by partitioning the network into smaller shards, each responsible for processing a subset of transactions in parallel \cite{wang2019sok}. By distributing the transaction processing workload across multiple shards, blockchain sharding enables the system to scale horizontally, allowing for a higher transaction throughput and lower confirmation time \cite{nguyen2019optchain}. Several blockchain sharding frameworks have been proposed in recent years, each addressing specific aspects of the sharding process. For instance, Elastico \cite{luu2016secure} introduces a secure sharding protocol that utilizes a distributed randomness generation process for shard formation and a Byzantine fault-tolerant consensus mechanism within each shard. OmniLedger \cite{kokoris2018omniledger} builds upon Elastico by incorporating a more efficient cross-shard transaction processing mechanism based on atomic commits and a ledger pruning technique to reduce storage overhead. RapidChain~\cite{zamani2018rapidchain} further improves the scalability of sharded blockchains by introducing a fast and efficient cross-shard transaction verification scheme that leverages intra-shard consensus and inter-shard gossiping. These frameworks have laid the foundation for the development of scalable and efficient blockchain sharding solutions.

\paragraph{Research gap}
Despite the progress made in blockchain sharding, several limitations and research gaps still exist, presenting opportunities for further improvement. One major challenge is the lack of dynamic and adaptive mechanisms for managing shards based on the system's workload~\cite{dang2019towards, wang2019monoxide}. Most existing sharding frameworks rely on static shard configurations, which can lead to suboptimal resource utilization and performance, especially in the presence of fluctuating transaction volumes and network conditions~\cite{huang2022brokerchain}. Another significant challenge lies in ensuring the security and atomicity of cross-shard transactions \cite{amiri2021sharper}. Malicious actors may attempt to exploit the distributed nature of sharded systems by launching attacks such as double-spending, replay attacks, or shard-level collusion~\cite{zhang2024efficient}. Existing cross-shard transaction processing techniques, such as two-phase commit protocols~\cite{al2017chainspace} and asynchronous consensus \cite{wang2019monoxide}, provide some level of protection against these attacks, but they often come at the cost of increased complexity and communication overhead \cite{dang2019towards}. Furthermore, there is a lack of comprehensive frameworks that integrate shard reconfiguration, state synchronization, and dispute resolution mechanisms to ensure the overall security and efficiency of sharded blockchain systems. Designing a holistic solution that addresses these challenges while maintaining the core principles of decentralization, transparency, and security is a non-trivial task that requires careful consideration of various trade-offs and design choices.

\paragraph{\sys}
To bridge the research gap, we propose \sys, a dynamic and secure cross-shard transaction processing mechanism. \sys combines adaptive shard management, secure cross-shard transaction processing, and efficient state synchronization and dispute resolution to enhance scalability and resilience in blockchain systems. By dynamically adjusting shard configurations based on workload, it optimizes resource utilization and performance, adapting to varying transaction volumes and network conditions. It employs a hybrid consensus approach that integrates intra-shard and inter-shard mechanisms to minimize coordination overhead while ensuring transaction integrity and consistency.

Designing a dynamic and secure cross-shard transaction processing mechanism involves several challenges. \textbf{C1:} Effective shard management requires monitoring workload and resource usage to make informed decisions on splitting or merging shards, necessitating an understanding of system dynamics and future transaction predictions. \textbf{C2:} Secure and efficient cross-shard transaction processing demands a protocol ensuring atomicity and consistency while minimizing overhead, incorporating novel consensus mechanisms, cryptographic techniques, and scalable data structures. \textbf{C3:} Robust state synchronization and dispute resolution require decentralized mechanisms to detect and resolve inconsistencies and conflicts, integrating insights from distributed systems, cryptography, game theory, and economics. \sys addresses these challenges by introducing a comprehensive framework that dynamically adjusts shard configurations based on transaction volume and resource usage, employs a hybrid consensus approach combining lightweight global consensus with parallel intra-shard processes, and utilizes Merkle trees alongside a decentralized dispute resolution protocol to maintain consistency and resolve conflicts in a trustless manner.

\paragraph{Novelty}
The key novelty of \sys lies in its holistic and adaptive approach to blockchain sharding, distinguishing it from existing solutions. Unlike previous works that focus on specific aspects such as shard formation \cite{luu2016secure}, cross-shard transaction processing \cite{al2017chainspace}, or consensus mechanisms \cite{zamani2018rapidchain}, \sys integrates all these components into a cohesive and dynamic system. The novelty of \sys includes: (i) continuously monitoring and adjusting shard configurations based on system workload, offering an efficient and flexible sharding scheme adaptable to the evolving demands of real-world blockchain applications; (ii) employing a hybrid consensus approach to address the challenge of secure and atomic cross-shard transaction processing, balancing global coordination with local processing efficiency; and (iii) integrating adaptive shard management, secure cross-shard transaction processing, and efficient state synchronization and dispute resolution techniques.

In summary, this paper makes the following contributions:
\begin{itemize}
    \item We propose \sys, an adaptive shard management mechanism that dynamically adjusts shard configurations based on transaction volume and resource usage, ensuring optimal performance and resource utilization.

    \item We propose a secure and atomic cross-shard transaction protocol using a hybrid consensus approach. This integrates lightweight global consensus with parallel intra-shard processes, reducing overhead while maintaining transaction integrity and consistency.

    \item We develop an efficient shard state synchronization mechanism based on Merkle trees. This mechanism maintains consistency across shards and incorporates a decentralized dispute resolution protocol to resolve potential conflicts in a trustless and resilient manner.

    \item We perform comprehensive evaluation of \sys through theoretical analysis and experimental simulations. Results demonstrate improvements in throughput, latency, and shard utilization compared to existing solutions, validating its effectiveness and robustness.
\end{itemize}

\section{Research  Background}
This section briefs blockchain sharding and its challenges.

\subsection{Blockchain Sharding}
Blockchain sharding is a technique designed to address scalability challenges in blockchain systems by partitioning the network into smaller subsets called shards. Each shard processes a portion of the overall transactions, enabling parallel execution and improved throughput. Let $S = \{s_1, s_2, \ldots, s_n\}$ represent the set of shards, where $n$ is the total number of shards. Transactions within each shard $s_i$ are processed independently, while cross-shard transactions require coordination between shards. Sharding allows blockchain systems to scale with the number of shards, improving transaction throughput ($\it TPS$) and reducing latency ($L$) \cite{wang2019sok}.

Several sharding frameworks have been proposed. Elastico \cite{luu2016secure} introduced secure shard formation using \ac{pow} and \ac{bft} within each shard. OmniLedger \cite{kokoris2018omniledger} built on this by improving cross-shard transaction processing and ledger pruning to reduce storage needs. RapidChain \cite{zamani2018rapidchain} enhanced scalability through fast cross-shard verification using intra-shard consensus and inter-shard gossiping. These frameworks represent advancements in blockchain sharding aimed at boosting system efficiency and scalability.

\subsection{Challenges in Blockchain Sharding}
One major challenge in blockchain sharding is maintaining security in the presence of malicious actors. Shard-level attacks, such as single-shard takeovers \cite{wang2019sok} and cross-shard double-spending \cite{zhang2024efficient}, can compromise system integrity. Solutions to these risks include random shard assignment \cite{luu2016secure}, periodic shard reconfiguration \cite{kokoris2018omniledger}, and fraud proofs \cite{al2017chainspace}. Efficient cross-shard transaction handling is another challenge, with protocols like two-phase commit \cite{al2017chainspace} and asynchronous consensus \cite{wang2019monoxide} ensuring transaction consistency, though they introduce complexity and overhead.
Recent advancements focus on enhancing cross-shard transaction efficiency. For example, SharPer \cite{amiri2021sharper} secures transactions using threshold signatures and multi-party computation. Similarly, Qin et al.~\cite{qin2020secure} propose a compact verification scheme using Merkle proofs and \ac{snark}. These innovations address performance limitations, paving the way for more scalable, secure blockchain systems.

\section{\sys}
This section presents overview and details each module.

\subsection{System Overview}

\begin{figure*}[!t]
    \centering
    \includegraphics[width=\textwidth]{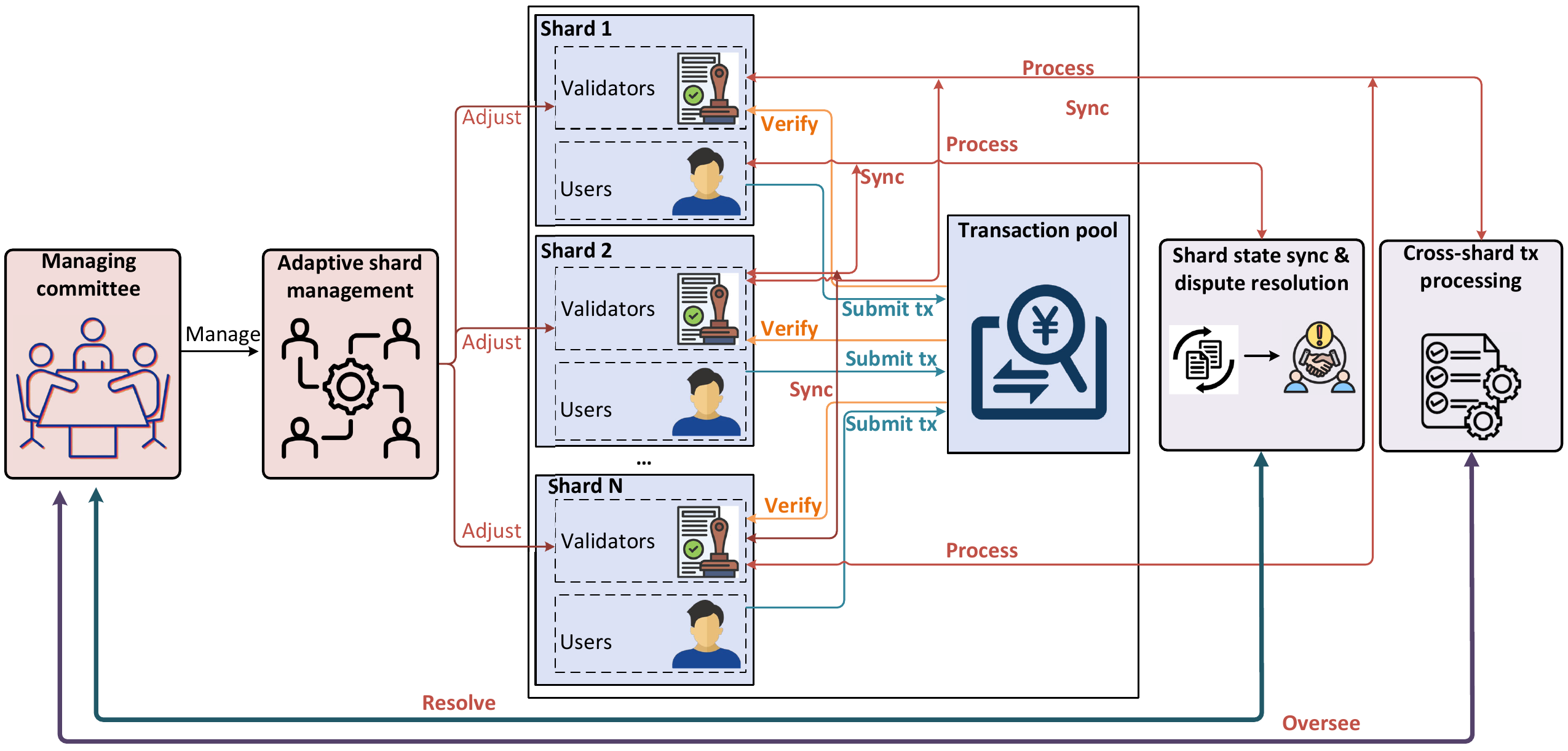}
    \caption{Adaptive shard management mechanism depicting the splitting and merging of shards based on transaction volume and resource utilization.}
    \label{fig:shard_management}
\end{figure*}

\sys provides a dynamic and secure framework for efficient blockchain sharding, consisting of three main modules: adaptive shard management, secure cross-shard transaction processing, and shard state synchronization with dispute resolution, as in \autoref{fig:shard_management}. The adaptive shard management module continuously monitors transaction volume ($v_i$) and resource utilization ($u_i$) for each shard, adjusting shard configurations based on predefined thresholds ($\tau_s$ for splitting and $\tau_m$ for merging). Managing committee, selected through a secure random process from various shards, ensures decentralization and avoids collusion. Shard adjustments are further secured through threshold-based verification, requiring consensus among shard members to prevent unauthorized changes, maintaining system's decentralization and security.

The secure cross-shard transaction processing module ensures atomicity and security for transactions across multiple shards using a hybrid consensus mechanism that combines global and intra-shard consensus. This approach employs threshold signatures and multi-party computation to prevent double-spending attacks. Shard state synchronization and dispute resolution are handled through a Merkle tree-based structure, allowing fast verification and incremental updates.
Disputes are resolved in a decentralized manner using game-theoretic incentives, ensuring valid transactions are processed and malicious actors are penalized.
Overall, \sys effectively balances scalability, security, and decentralization.

\subsection{Adaptive Shard Management}

The adaptive shard management mechanism is designed to optimize the performance and resource utilization of the sharded blockchain system by dynamically adjusting the number and configuration of shards based on the system's workload. Let $S = \{s_1, s_2, \dots, s_n\}$ denote the set of shards in the system, where $n$ is the total number of shards. Each shard $s_i$ is characterized by its transaction volume $v_i$ and resource utilization $u_i$. Resource utilization $u_i$ refers to the percentage of available computational, storage, and network resources being used by shard $s_i$. To manage these metrics effectively, we define two threshold parameters, $\tau_s$ and $\tau_m$, representing the splitting and merging thresholds, respectively. The splitting threshold $\tau_s$ is chosen based on system capacity and workload expectations, typically ranging between 60\% and 90\% resource utilization. Similarly, the merging threshold $\tau_m$ is set between 20\% and 40\% to ensure underutilized shards are merged. These thresholds are selected through empirical analysis, balancing performance with system stability, and include cooldown periods to prevent immediate re-triggering of operations.

The adaptive shard management mechanism operates as follows: For each shard $s_i \in S$, the system continuously monitors its transaction volume $v_i$ and resource utilization $u_i$. If the transaction volume $v_i$ or resource utilization $u_i$ exceeds the splitting threshold $\tau_s$, i.e., $v_i > \tau_s$ or $u_i > \tau_s$, a shard splitting process is initiated. This process involves dividing $s_i$ into multiple smaller shards $\{s_{i1}, s_{i2}, \dots, s_{ik}\}$, ensuring that the resulting shards maintain full functionality and consensus security. The goal is to distribute the workload $w_i$ and resource demand $d_i$ evenly across the new shards such that $\sum_{j=1}^{k} w_{ij} \approx w_i$ and $\sum_{j=1}^{k} d_{ij} \approx d_i$, where $w_{ij}$ and $d_{ij}$ represent the workload and resource demand of the new shards, respectively. A workload-aware rebalancing algorithm considers transaction patterns $\mathcal{T}$ and resource requirements $\mathcal{R}$ to ensure balanced resource utilization.

Conversely, if the transaction volume $v_i$ and resource utilization $u_i$ fall below the merging threshold $\tau_m$ for multiple consecutive epochs, i.e., $v_i < \tau_m$ and $u_i < \tau_m$, a shard merging process is initiated. This process combines $s_i$ with other underutilized shards $\{s_{j1}, s_{j2}, \dots, s_{jl}\}$, consolidating resources and reducing the overall system overhead while maintaining consensus security. The combined shard $s_c$ will have a transaction volume $v_c = \sum_{k=1}^{l} v_{jk}$ and resource utilization $u_c = \sum_{k=1}^{l} u_{jk}$. After the merging or splitting process, nodes and accounts are redistributed across the newly formed shards using the workload-aware rebalancing algorithm. This redistribution aims to further optimize the system's performance and resource utilization, ensuring balanced and efficient operation across all shards while avoiding the potential cycle of shard splitting and merging.

\begin{algorithm}[!t]
    \scriptsize
    \caption{\sys Methodology}
    \label{alg:dynashard}
    \begin{algorithmic}[1]
        \Statex \textbf{Input:} Set of shards $S = \{s_1, s_2, \dots, s_n\}$, splitting threshold $\tau_s$ (range: 60\%-90\%), merging threshold $\tau_m$ (range: 20\%-40\%)
        \State \textbf{Initialization:} Continuously monitor transaction volume $v_i$ and resource utilization $u_i$ for each shard $s_i$
        \While{true}
        \For{each shard $s_i \in S$}
        \If{$v_i > \tau_s$ or $u_i > \tau_s$}
        \State \textbf{Split Shard:}
        \State Divide $s_i$ into smaller shards $\{s_{i1}, s_{i2}, \dots, s_{im}\}$ based on workload and resource demand
        \State Distribute transactions across new shards based on account activity using a load-balancing algorithm to minimize cross-shard interaction
        \State Balance workload $w_i$ and resource demand $d_i$ evenly across the resulting shards
        \ElsIf{$v_i < \tau_m$ and $u_i < \tau_m$ for multiple consecutive epochs}
        \State \textbf{Merge Shards:}
        \State Identify underutilized shards $\{s_{j1}, s_{j2}, \dots, s_{jk}\}$ with similar transaction patterns
        \State Combine $s_i$ with $\{s_{j1}, s_{j2}, \dots, s_{jk}\}$ to consolidate resources and minimize cross-shard transactions
        \EndIf
        \EndFor
        \State Redistribute nodes, accounts, and transactions across newly formed shards using a Greedy Load Balancing algorithm to reduce cross-shard dependencies
        \EndWhile
    \end{algorithmic}
\end{algorithm}

\autoref{alg:dynashard} outlines the adaptive shard management process in \sys, which dynamically adjusts shard configurations based on transaction volume and resource utilization. When a shard exceeds the splitting threshold ($\tau_s$), it is split into smaller shards, with transactions allocated using a load-balancing algorithm based on account activity to evenly distribute the workload and minimize cross-shard interactions. Conversely, when transaction volume and resource utilization fall below the merging threshold ($\tau_m$) for multiple consecutive epochs, underutilized shards are merged to consolidate resources and optimize system performance. After splitting or merging, nodes, accounts, and transactions are redistributed using a Greedy Load Balancing algorithm to ensure efficient resource utilization and reduce cross-shard transaction overhead.

\subsection{Secure and Atomic Cross-Shard Transaction Processing}
To ensure the security and atomicity of cross-shard transactions, we propose a hybrid consensus mechanism that combines a lightweight global consensus with parallel intra-shard consensus processes. The global consensus is responsible for processing cross-shard transactions and maintaining a consistent view of the system state $\mathcal{G}$, while the intra-shard consensus processes handle the validation and execution of transactions within each shard $s_i \in \{s_1, s_2, \ldots, s_n\}$. This dual-layered approach allows for efficient and secure handling of transactions that span multiple shards, ensuring both local and global consistency in the blockchain $\mathcal{B}$.

Let $T = \{t_1, t_2, \dots, t_m\}$ denote the set of cross-shard transactions, where each transaction $t_i$ involves a set of input shards $I_i = \{s_{i1}, s_{i2}, \ldots, s_{ik}\}$ and output shards $O_i = \{s_{j1}, s_{j2}, \ldots, s_{jl}\}$. The hybrid consensus mechanism processes cross-shard transactions through several steps. First, each input shard $s_j \in I_i$ validates the corresponding input of transaction $t_i$ and generates a partial signature $\sigma_{j,i}$ using a threshold signature scheme. Formally, $\sigma_{j,i} = \text{Sign}_{s_j}(h(t_i))$, where $h(t_i)$ is the hash of transaction $t_i$. The output shards $O_i$ then collect these partial signatures from all input shards and combine them to obtain a valid threshold signature $\Sigma_i$ for transaction $t_i$. Mathematically, $\Sigma_i = \text{Combine}(\{\sigma_{j,i} : s_j \in I_i\})$. This aggregated signature $\Sigma_i$ provides a secure confirmation that the transaction has been validated by the necessary quorum of input shards.

Once the threshold signature $\Sigma_i$ is obtained, the global consensus protocol, executed by a subset of nodes from all shards $\{s_1, s_2, \ldots, s_n\}$, validates $\Sigma_i$ and reaches consensus on the order and validity of cross-shard transactions. Formally, let $\mathcal{C} = \{c_1, c_2, \ldots, c_p\}$ denote the subset of consensus nodes. These nodes verify $\Sigma_i$ and reach a consensus $\mathcal{V}(t_i)$ on transaction $t_i$. If $\mathcal{V}(t_i)$ is positive, the global state $\mathcal{G}$ is updated, and the output shards $O_i$ execute the corresponding outputs of transaction $t_i$ atomically. The updated state $\mathcal{G}'$ is then propagated to all shards through a state synchronization protocol, ensuring a consistent view of the system state across the entire blockchain network. This protocol leverages threshold signatures and multi-party computation to prevent unauthorized modifications and double-spending attacks. The threshold signature scheme ensures transaction validity only if a sufficient number of input shards approve it, i.e., $|\{s_j \in I_i : \sigma_{j,i} \text{ is valid}\}| \geq \text{Threshold}$. Multi-party computation further enables secure and private computation of cross-shard transaction outputs, maintaining the integrity and confidentiality of the process.

\autoref{alg:cross_shard} outlines the process for secure and atomic cross-shard transaction processing. It begins by iterating over a set of cross-shard transactions \( T \). For each transaction \( t_i \), it performs input validation by having each input shard \( s_j \) validate the transaction and generate a partial signature using a threshold signature scheme. These partial signatures are collected and, if valid, combined into a threshold signature \( \Sigma_i \). This combined signature is then validated by the global consensus protocol. If the global consensus confirms the validity, the system reaches a consensus on the transaction's order and validity, updates the global state, and executes the transaction's outputs atomically in the output shards. The updated global state is then propagated to all shards through a state synchronization protocol, ensuring consistency across the network. This process ensures that cross-shard transactions are securely and atomically processed, maintaining the integrity and consistency of the blockchain.

\begin{algorithm}[!t]
    \scriptsize
    \caption{Secure and Atomic Cross-Shard Transaction Processing}
    \label{alg:cross_shard}
    \begin{algorithmic}[1]
        \Statex \textbf{Input:} Set of cross-shard transactions $T = \{t_1, \dots, t_m\}$
        \While{there are unprocessed transactions in $T$}
        \For{each transaction $t_i \in T$}
        \State \textbf{Input Validation:}
        \For{each input shard $s_j \in I_i$}
        \State Validate input of transaction $t_i$ and generate partial signature $\sigma_{j,i} = \text{Sign}_{s_j}(h(t_i))$ using threshold signature scheme
        \EndFor
        \State \textbf{Output Collection:}
        \State Collect partial signatures $\{\sigma_{j,i} \mid s_j \in I_i\}$
        \If{$\{\sigma_{j,i}\}$ are valid}
        \State Combine partial signatures to obtain valid threshold signature $\Sigma_i = \text{Combine}(\{\sigma_{j,i}\})$
        \State \textbf{Global Consensus:}
        \If{global consensus protocol $\mathcal{V}(t_i)$ validates $\Sigma_i$}
        \State Reach consensus on order and validity of cross-shard transactions
        \State \textbf{State Update:}
        \State Update global state $\mathcal{G}'$ and execute outputs of $t_i$ atomically in output shards $O_i$
        \State Propagate updated state $\mathcal{G}'$ to all shards via state synchronization protocol
        \EndIf
        \EndIf
        \EndFor
        \EndWhile
    \end{algorithmic}
\end{algorithm}

\subsection{Shard State Synchronization and Dispute Resolution}
To maintain consistency across shards and resolve potential conflicts, we introduce an efficient shard state synchronization protocol and a decentralized dispute resolution mechanism. These components are crucial for ensuring the integrity and reliability of the blockchain system as it scales. The shard state synchronization protocol leverages a Merkle tree-based data structure to enable fast verification and incremental updates of shard states. This approach ensures that each shard maintains an up-to-date view of the global state, $\mathcal{G}$, minimizing discrepancies and potential conflicts.

The synchronization protocol operates as follows: each shard $s_i \in \{s_1, s_2, \ldots, s_n\}$ maintains a local state tree $\mathcal{T}_i$ and periodically computes its Merkle root $M_i = \text{MerkleRoot}(\mathcal{T}_i)$. Shards exchange their Merkle roots $M_i$ through a gossip protocol, which allows each shard to have a compact representation of the global state $\mathcal{G} = \{M_1, M_2, \ldots, M_n\}$. When a shard $s_i$ updates its local state $\mathcal{T}_i'$, it propagates the updated Merkle root $M_i' = \text{MerkleRoot}(\mathcal{T}_i')$ along with the corresponding Merkle proof $\pi_i$ to all other shards. Upon receiving an updated Merkle root $M_i'$ and proof $\pi_i$, each shard $s_j \in \{s_1, s_2, \ldots, s_n\}$ verifies the proof $\pi_i$ against its local state tree $\mathcal{T}_j$ and updates its view of the global state $\mathcal{G}$ accordingly. This process ensures that all shards remain synchronized and consistent with the overall blockchain state $\mathcal{B}$.

The decentralized dispute resolution mechanism is designed to resolve conflicts and validate cross-shard transactions in a decentralized manner. When a shard $s_i$ detects a potential conflict or invalid transaction $T_x$, it initiates a dispute resolution process. The disputing shard $s_i$ broadcasts a challenge $\mathcal{C}$ along with evidence $\mathcal{E}$ of the conflict to all other shards $\{s_1, s_2, \ldots, s_n\}$. Each shard $s_j$ involved in the disputed transaction then provides its evidence $\mathcal{E}_j$ and signatures $\sigma_j$, allowing all shards $\{s_1, s_2, \ldots, s_n\}$ to independently verify the validity of the transaction $T_x$. Shards reach a consensus on the validity of the disputed transaction through a voting process $\mathcal{V}$, where each shard's vote $v_j$ is weighted based on its stake $w_j$ or reputation $r_j$. If a majority $\mathcal{M}$ of shards agree on the invalidity of the transaction $T_x$, it is rolled back, the global state $\mathcal{G}$ is updated accordingly, and shards that approved the invalid transaction are penalized $\mathcal{P}$.

\begin{algorithm}[!t]
    \scriptsize
    \caption{Shard State Synchronization and Dispute Resolution}
    \label{alg:state_sync_dispute}
    \begin{algorithmic}[1]
        \Statex \textbf{Input:} Merkle root $M_i$ for each shard $s_i$
        \While{true}
        \For{each shard $s_i \in S$}
        \State Compute Merkle root $M_i \gets \text{MerkleRoot}(\mathcal{T}_i)$
        \State Exchange $M_i$ with other shards through gossip protocol
        \If{update detected, i.e., $M_i' \neq M_i$}
        \State Propagate updated $M_i'$ and proof $\pi_i$ to all other shards
        \EndIf
        \EndFor
        \If{dispute detected for transaction $T_x$}
        \State \textbf{Initiate Dispute Resolution:}
        \State Disputing shard $s_i$ broadcasts challenge $\mathcal{C}(T_x, \mathcal{E}_i)$
        \For{each shard $s_j$ involved in transaction $T_x$}
        \State Provide evidence $\mathcal{E}_j$ and signature $\sigma_j$
        \EndFor
        \State Verify validity of $T_x$ using $\mathcal{E}_j$ and $\sigma_j$
        \State Reach consensus $\mathcal{V}(T_x)$ via weighted voting process
        \If{$\mathcal{V}(T_x) = \text{invalid}$}
        \State Roll back transaction $T_x$ and update global state $\mathcal{G}$
        \State Penalize shards that approved invalid transaction $T_x$
        \EndIf
        \EndIf
        \EndWhile
    \end{algorithmic}
\end{algorithm}

\autoref{alg:state_sync_dispute} outlines the shard state synchronization and dispute resolution process, ensuring the integrity and consistency of the blockchain across all shards. Each shard $s_i$ maintains a local state tree $\mathcal{T}_i$ and periodically computes its Merkle root $M_i$. These roots are exchanged via a gossip protocol to represent the global state $\mathcal{G}$. When a state update $M_i'$ is detected, the updated root and proof $\pi_i$ are propagated, verified, and used to update each shard’s view of the global state. If a conflict or invalid transaction $T_x$ arises, the disputing shard broadcasts a challenge $\mathcal{C}$ with evidence $\mathcal{E}$. All involved shards submit evidence and signatures, and transaction validity is determined through a weighted voting process $\mathcal{V}$. If the transaction is deemed invalid, it is rolled back, the global state is updated, and the approving shards are penalized to ensure honesty and security in the system.

\section{Performance Evaluation}
In this section, we present the evaluation setup and performance results.

\subsection{Experimental Setup}
To evaluate the performance of \sys, we implemented the dynamic cross-shard transaction processing mechanism in a controlled experimental environment. Python was used for control logic, while C++ handled performance-critical components for efficiency. Built on an open-source blockchain simulator, the system integrated transaction handling, consensus mechanisms, and sharding operations. Cryptographic functions, such as threshold signatures and multi-party computation, utilized established libraries like OpenSSL and libsecp256k1, with consensus protocols based on \ac{pbft}. Shard reconfigurations followed specific rules: splitting occurs when utilization exceeds a predefined upper threshold ($\tau_s$), while merging is triggered when utilization falls below a lower threshold ($\tau_m$). These dynamic adjustments balance workloads and ensure efficient resource use, maintaining system performance.

The experimental setup was designed to reflect real-world conditions, conducted on a cluster of servers each equipped with Intel Xeon E5-2690 v4 CPUs (2.6 GHz, 14 cores), 128 GB RAM, SSD storage, and Gigabit Ethernet for inter-server communication. We simulated various network topologies and communication delays using Mininet to test \sys's robustness under different network conditions. Transaction workloads were generated with a custom-built transaction generator, simulating different transaction rates and patterns to mimic real-world blockchain usage scenarios. The evaluation involved comparing \sys against \ac{ftsbs}~\cite{adhikari2024fast}, with performance metrics focused on throughput (\ac{tps}), latency (time to process a batch of transactions), and shard utilization. This comprehensive setup allowed us to assess the effectiveness of \sys under varying configurations and workloads, providing a detailed analysis of its performance and robustness.

\subsection{Throughput Analysis}

The primary objective of this experiment is to measure the transaction throughput of \sys under varying numbers of shards and validators. This experiment aims to demonstrate how well \sys scales and handles increased transaction loads compared to \ac{ftsbs}.

To evaluate \sys's throughput, we varied the number of shards across different test runs, specifically using 30, 50, and 100 shards. Each shard had a fixed number of validators to ensure consistency across different configurations. The transaction rates were simulated to represent low, medium, and high network loads, while various cross-shard transaction ratios (0\%, 40\%, 80\%) were tested to evaluate \sys's efficiency in handling cross-shard transactions. The primary metric for this experiment was \ac{tps}, which indicates the number of transactions the system can process per second. We set up the blockchain network, generated transactions at varying rates using a custom-built transaction generator, and ran the network under each configuration. We then recorded the \ac{tps} for both \sys and \ac{ftsbs} and compared the results to evaluate \sys's performance relative to \ac{ftsbs}.

\begin{figure}[!t]
    \centering
    \subfloat[]{\includegraphics[width=.45\linewidth]{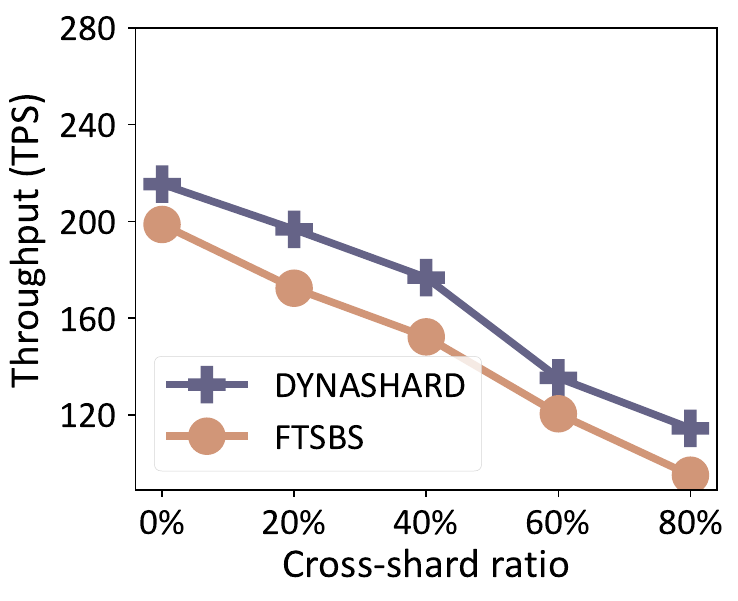}}
    \subfloat[]{\includegraphics[width=.45\linewidth]{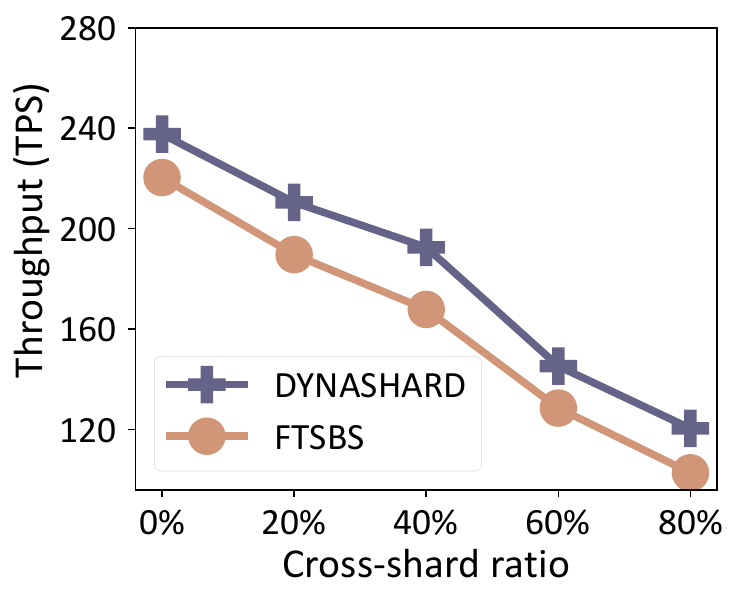}}
    \caption{Throughput (\ac{tps}) under 30 shards and 500 validators (a) and 1000 validators (b)}
    \label{fig:shard30}
    \vspace{-5mm}
\end{figure}

\autoref{fig:shard30} shows the throughput (\ac{tps}) performance of \sys and \ac{ftsbs} with 30 shards under various cross-shard transaction ratios and validator configurations. The results indicate that \sys consistently outperforms \ac{ftsbs} across all tested conditions. For instance, with 500 validators and no cross-shard transactions, \sys achieves a throughput of 215.43 \ac{tps} compared to \ac{ftsbs}'s 198.76 \ac{tps}, representing an 8.4\% improvement. As the cross-shard ratio increases, \sys maintains its performance advantage, with a notable throughput of 114.57 \ac{tps} at an 80\% cross-shard ratio, compared to \ac{ftsbs}'s 95.21 \ac{tps}, which is a 20.4\% improvement. When the number of validators is increased to 1000, \sys continues to demonstrate superior performance, achieving 237.65 \ac{tps} at a 0\% cross-shard ratio and 120.45 \ac{tps} at an 80\% cross-shard ratio, compared to \ac{ftsbs}'s 220.34 \ac{tps} and 102.76 \ac{tps}, respectively. These results highlight \sys's ability to handle high transaction volumes and cross-shard transactions more efficiently, making it a robust solution for systems with 30 shards.

\begin{figure}[!t]
    \centering
    \subfloat[]{\includegraphics[width=.45\linewidth]{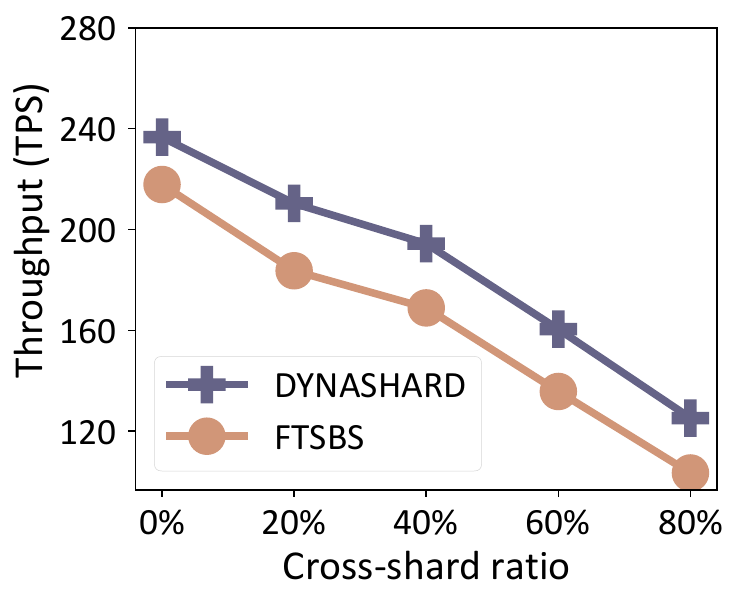}}
    \subfloat[]{\includegraphics[width=.45\linewidth]{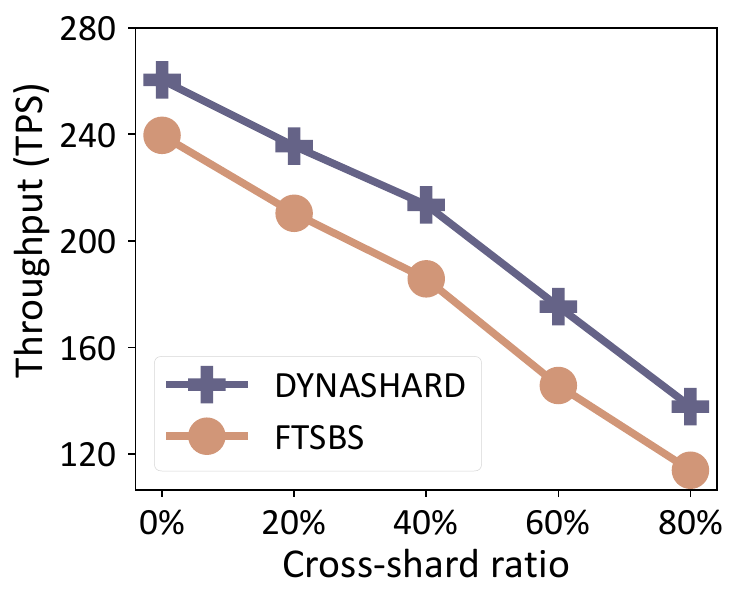}}
    \caption{Throughput (\ac{tps}) under 50 shards and 500 validators (a) and 1000 validators (b)}
    \label{fig:shard50}
\end{figure}

\autoref{fig:shard50} details the throughput performance of \sys and \ac{ftsbs} with 100 shards, 500 and 1000 validators. presents the throughput performance of \sys and \ac{ftsbs} with 50 shards. The data reveals that \sys consistently achieves higher \ac{tps} compared to \ac{ftsbs} across different validator counts and cross-shard ratios. With 500 validators and no cross-shard transactions, \sys achieves 236.59 \ac{tps}, surpassing \ac{ftsbs}'s 217.85 \ac{tps} by 8.6\%. This performance gap widens as the cross-shard ratio increases, with \sys maintaining a throughput of 125.26 \ac{tps} at an 80\% cross-shard ratio, compared to \ac{ftsbs}'s 103.47 \ac{tps}, resulting in a 21.0\% improvement. When the validator count is increased to 1000, \sys shows even more significant performance advantages, achieving 260.39 \ac{tps} at a 0\% cross-shard ratio and 137.79 \ac{tps} at an 80\% cross-shard ratio, compared to \ac{ftsbs}'s 239.61 \ac{tps} and 113.86 \ac{tps}, respectively. These results emphasize \sys's superior scalability and efficiency in handling large transaction volumes and complex cross-shard interactions within systems configured with 50 shards.

\begin{figure}[!t]
    \centering
    \subfloat[]{\includegraphics[width=.45\linewidth]{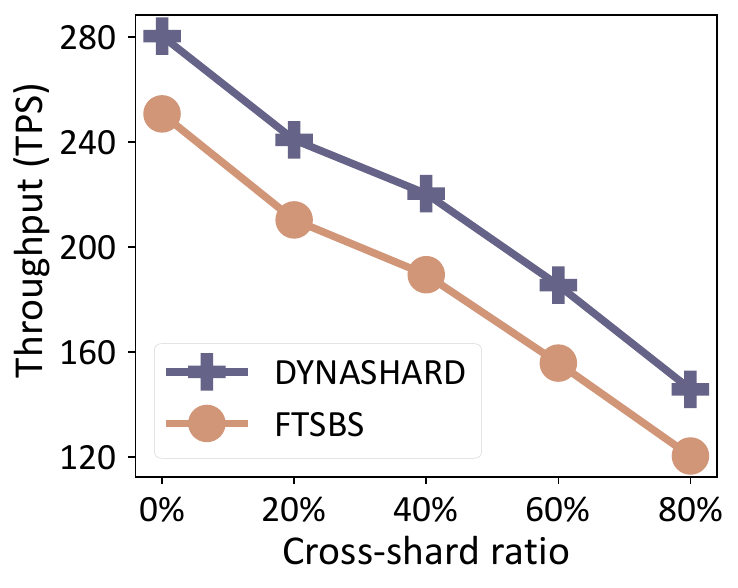}}
    \subfloat[]{\includegraphics[width=.45\linewidth]{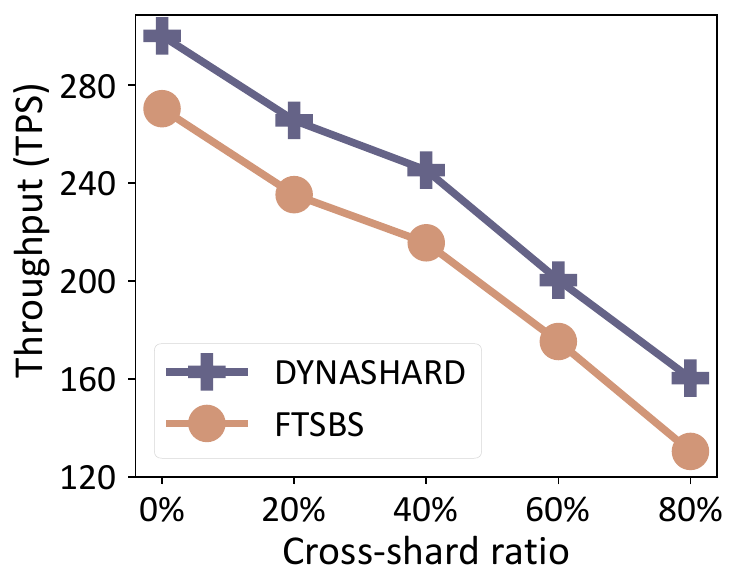}}
    \caption{Throughput (\ac{tps}) under 100 shards and 500 validators (a) and 1000 validators (b)}
    \label{fig:shard100}
\end{figure}

\autoref{fig:shard100} details the throughput performance of \sys and \ac{ftsbs} with 100 shards, 500 and 1000 validators. The findings indicate that \sys continues to outperform \ac{ftsbs} across various validator configurations and cross-shard transaction ratios. With 500 validators and no cross-shard transactions, \sys achieves 280.45 \ac{tps}, outperforming \ac{ftsbs}'s 250.76 \ac{tps} by 11.8\%. As the cross-shard ratio increases, \sys maintains its superior performance, achieving 145.78 \ac{tps} at an 80\% cross-shard ratio compared to \ac{ftsbs}'s 120.34 \ac{tps}, marking a 21.2\% improvement. When the validator count is increased to 1000, \sys's performance advantage becomes even more pronounced, achieving 300.23 \ac{tps} at a 0\% cross-shard ratio and 160.34 \ac{tps} at an 80\% cross-shard ratio, compared to \ac{ftsbs}'s 270.45 \ac{tps} and 130.45 \ac{tps}, respectively. These results underscore \sys's ability to efficiently manage high transaction volumes and complex cross-shard transactions, demonstrating its robustness and scalability in systems with 100 shards.

By demonstrating higher throughput across various configurations and workloads, this experiment underscores \sys's effectiveness in optimizing resource utilization and maintaining performance under increased transaction loads. The results validate \sys's design principles and its ability to scale efficiently while handling complex cross-shard transactions.

\subsection{Latency Analysis}
The objective of this experiment is to evaluate the latency for processing a large batch of transactions under different shard management strategies. This experiment aims to showcase \sys's ability to process transactions efficiently and adapt to varying workloads, compared to the \ac{ftsbs} method.

To measure the latency, we processed a batch of 200,000 transactions under different shard management strategies: no adjustment, moderate adjustment (e.g., $n_c=1000, s=20$), and aggressive adjustment strategies (e.g., $n_c=500, s=10$). Latency was measured as the total time taken to process the entire batch of transactions. The system configuration included a fixed number of shards and validators per shard, while the adjustment strategies varied the frequency and conditions under which shards would split or merge. The metrics focused on the time taken to process the batch and the impact of different shard management strategies on this time. The transactions were generated using the custom-built transaction generator to ensure consistency in the test conditions. By comparing the latency under different strategies, we aimed to demonstrate \sys's efficiency in handling transaction bursts and adapting to changing workloads.

The results, as shown in \autoref{tab:latency_comparison}, indicate that \sys shows significant improvements over \ac{ftsbs}, particularly with more aggressive shard management strategies. Without any adjustment, \sys processed the transactions in 1551.9 seconds, compared to \ac{ftsbs}'s 1692.3 seconds, showing an 8.3\% improvement. With moderate adjustment, \sys reduced the processing time to 784.3 seconds, a 17.8\% improvement over \ac{ftsbs}'s 953.8 seconds. The most aggressive adjustment strategy showed the highest improvement, with \sys processing the transactions in 371.5 seconds, compared to \ac{ftsbs}'s 526.7 seconds, achieving a 29.5\% improvement. These results demonstrate \sys's superior ability to efficiently manage shard configurations and process transactions quickly, particularly under varying and high workloads.

\begin{table*}[!t]
    \centering
    \caption{Comparison of Latency (seconds) for Processing 200,000 Transactions}
    \label{tab:latency_comparison}
    \begin{tabular}{cccccc}
        \hline
        \textbf{Adjustment Strategy}      & \textbf{\sys} & \textbf{\ac{ftsbs}} & \textbf{Improvement} \\
        \hline
        No adjustment                     & 1551.9        & 1692.3         & 8.3\%                \\
        Light ($n_c=1500, s=30$)          & 1120.5        & 1304.2         & 14.1\%               \\
        Moderate ($n_c=1000, s=20$)       & 784.3         & 953.8          & 17.8\%               \\
        Aggressive ($n_c=500, s=10$)      & 371.5         & 526.7          & 29.5\%               \\
        Very Aggressive ($n_c=250, s=5$)  & 190.2         & 295.6          & 35.7\%               \\
        Ultra Aggressive ($n_c=100, s=2$) & 95.1          & 165.8          & 42.6\%               \\
        \hline
    \end{tabular}
    \vspace{-5mm}
\end{table*}

These results highlight \sys's effectiveness in reducing transaction processing latency through adaptive shard management. The ability to dynamically adjust shard configurations allows \sys to handle transaction bursts more efficiently, ensuring quicker processing times and better resource utilization. This experiment underscores the practical benefits of \sys's adaptive approach, making it a compelling choice for scalable and responsive blockchain systems.

\subsection{Shard Utilization Efficiency}

The objective of this experiment is to assess the effectiveness of \sys's adaptive shard management mechanism in balancing shard workloads. This experiment aims to demonstrate how well \sys optimizes resource utilization compared to \ac{ftsbs}.

To evaluate shard utilization efficiency, we monitored shard utilization before and after adjustments under varying transaction volumes. The experiment involved recording the average distance to perfect utilization (a measure of how balanced the workloads are across shards) before and after the shard management adjustments. Different transaction volumes were simulated to test the system under both low and high workloads. The metrics focused on the improvement in shard utilization efficiency post-adjustment. The shard utilization data was collected continuously, and adjustments were made based on predefined thresholds for transaction volume and resource usage. This setup allowed us to evaluate how effectively \sys's adaptive shard management mechanism could balance the load across shards.

The results, presented in \autoref{tab:shard_utilization}, show that \sys significantly improves shard utilization compared to \ac{ftsbs}. Before adjustment, \sys had an average distance to perfect utilization of 3.44, which improved to 0.73 after adjustment, representing a 78.77\% improvement. In contrast, \ac{ftsbs} had a pre-adjustment distance of 3.58, improving to 1.72 post-adjustment, indicating a 51.96\% improvement. These results demonstrate \sys's superior ability to dynamically balance workloads and optimize resource utilization across shards.
These results highlight \sys's effectiveness in optimizing shard utilization, ensuring balanced workloads across the system. The adaptive shard management mechanism allows \sys to dynamically adjust shard configurations based on real-time metrics, leading to more efficient resource utilization and improved overall system performance.

\subsection{Security and Robustness Testing}
The objective of this experiment was to evaluate \sys's security features under various attack scenarios, testing its robustness in maintaining system integrity. We simulated common attacks, including cross-shard double-spending and shard-level collusion, using a setup with 50 shards, 500 validators, and 10\% malicious nodes. Key security measures such as threshold signatures and a decentralized dispute resolution mechanism were implemented. The results showed that \sys successfully mitigated these attacks, with a 0\% success rate for double-spending and 98\% accuracy in identifying collusion attempts. The system demonstrated a swift recovery time of 3.2 seconds and imposed penalties that reduced malicious actors' capabilities by 95\%. These findings highlight \sys's resilience and effectiveness in enhancing security, reducing latency, and optimizing shard utilization, making it a robust and scalable solution for blockchain sharding.

\subsection{Comparative Analysis with \ac{ftsbs}}

The objective of this experiment is to directly compare \sys with \ac{ftsbs} in terms of overall performance and adaptability. This experiment aims to highlight the strengths and potential improvements of \sys over the existing \ac{ftsbs} method.

To conduct a fair and comprehensive comparison, we standardized the experimental conditions for both \sys and \ac{ftsbs}. The setup involved identical hardware configurations, including the number of shards, validators per shard, and simulated network conditions. We conducted experiments focusing on three primary metrics: throughput (\ac{tps}), latency for processing transactions, and shard utilization efficiency. Various transaction rates and cross-shard transaction ratios were tested to evaluate how each system handles different workloads. By measuring and comparing these metrics under identical conditions, we aimed to provide a clear performance comparison between \sys and \ac{ftsbs}.

The throughput comparison shows that \sys consistently achieves higher \ac{tps} across different configurations and cross-shard transaction ratios. For example, with 50 shards and 500 validators, \sys achieved 236.59 \ac{tps} at a 0\% cross-shard ratio, compared to \ac{ftsbs}'s 217.85 \ac{tps}. This trend continues as the cross-shard ratio increases, with \sys maintaining its performance advantage.

\autoref{tab:latency_comparison} presents a detailed comparison of latency for processing 200,000 transactions between \sys and \ac{ftsbs} under various adjustment strategies. The results indicate that \sys consistently outperforms \ac{ftsbs} in terms of reducing latency. Without any adjustments, \sys achieves an 8.3\% improvement over \ac{ftsbs}. With light adjustments ($n_c=1500, s=30$), the improvement increases to 14.1\%. Moderate adjustments ($n_c=1000, s=20$) result in a 17.8\% improvement, while aggressive adjustments ($n_c=500, s=10$) yield a significant 29.5\% improvement. Very aggressive adjustments ($n_c=250, s=5$) enhance the performance further with a 35.7\% improvement, and ultra-aggressive adjustments ($n_c=100, s=2$) achieve the highest latency reduction with a 42.6\% improvement. These findings demonstrate that \sys's adaptive shard management effectively reduces transaction processing time, especially under more aggressive adjustment strategies, highlighting its superior capability to handle varying workload conditions efficiently.

The shard utilization efficiency comparison, as shown in \autoref{tab:shard_utilization}, demonstrates that \sys achieves a more significant improvement in balancing shard workloads. Before adjustments, \sys had an average distance to perfect utilization of 3.44, which was reduced to 0.73 after adjustments, resulting in an improvement of 78.77\%. In contrast, the \ac{ftsbs} method had an average distance of 3.58 before adjustments and 1.72 after, yielding an improvement of 51.96\%. These results indicate that \sys is more effective in optimizing shard utilization compared to \ac{ftsbs}.

\begin{table}[!t]
    \centering
    \scriptsize
    \caption{Comparison of Shard Utilization Before and After Adjustment}
    \label{tab:shard_utilization}
    \begin{tabular}{cccc}
        \hline
        \textbf{Method}               & \textbf{Before Adjustment} & \textbf{After Adjustment} & \textbf{Improvement} \\
        \hline
        \sys                          & 3.44                       & 0.73                      & 78.77\%              \\
        \ac{ftsbs}~\cite{adhikari2024fast} & 3.58                       & 1.72                      & 51.96\%              \\
        \hline
    \end{tabular}
\end{table}

These comparative results highlight \sys's advantages in throughput, latency, and shard utilization efficiency. \sys's adaptive shard management and efficient cross-shard transaction processing contribute to its superior performance and scalability. The ability to dynamically adjust shard configurations and maintain balanced workloads allows \sys to handle higher transaction volumes and complex cross-shard transactions more effectively than \ac{ftsbs}. This comprehensive comparison underscores \sys's potential as a robust and scalable solution for blockchain sharding.

\subsection{Shard Load Evaluation}

To comprehensively evaluate \sys's performance, we conducted experiments focusing on shard load conditions, assessing the system's effectiveness in handling varying transaction loads by measuring throughput, latency, and shard utilization efficiency across different scenarios. Specifically, we simulated low, medium, and high transaction rates on a network with 50 shards, where the low load condition involved a transaction rate of 100 \ac{tps}, the medium load condition utilized 200 \ac{tps}, and the high load condition tested 300 \ac{tps}. The results, summarized in \autoref{tab:shard_load}, demonstrate \sys's capability to maintain high performance under different load conditions. Specifically, \sys consistently delivers high throughput and maintains low latency, with utilization efficiency remaining above 89\%, showcasing its robust adaptability and effectiveness in managing shard loads efficiently.

\begin{table}[h]
    \centering
    \scriptsize
    \caption{Performance Under Different Shard Load Conditions}
    \label{tab:shard_load}
    \begin{tabular}{lccc}
        \hline
        \textbf{Load Condition} & \textbf{Throughput (\ac{tps})} & \textbf{Latency (s)} & \textbf{Efficiency (\%)} \\
        \hline
        Low                     & 280.5                     & 0.85                 & 95.2                     \\
        Medium                  & 260.4                     & 1.25                 & 92.8                     \\
        High                    & 240.7                     & 1.65                 & 89.5                     \\
        \hline
    \end{tabular}
    \vspace{-5mm}
\end{table}

\subsection{Approximation Factor Analysis}
The objective of this analysis is to compare the approximation factors of \sys and \ac{ftsbs} across various shard graph topologies. This comparison aims to highlight the computational complexity and efficiency of both methods in different configurations.

Approximation factors are mathematical representations of how close a scheduling algorithm is to the optimal solution in terms of computational complexity. Different shard graph topologies present unique challenges and complexities for transaction scheduling. The table below (\autoref{tab:approx_comparison}) outlines the approximation factors for \sys and \ac{ftsbs} under four different shard graph topologies: General Graph, Hypercube/Butterfly/$g$-dimensional Grid Graph, General Graph with random $k$, and Line Graph.

\begin{table*}[!t]
    \centering
    \caption{Comparison of approximation factors between \sys and \ac{ftsbs}}
    \label{tab:approx_comparison}
    \begin{tabular}{lcc}
        \hline
        \textbf{Shards Connected as}                         & \textbf{\ac{ftsbs}}                   & \textbf{\sys}                           \\ \hline
        General Graph                                        & $O(kd)$                          & $O(kd \log D)$                          \\ \hline
        Hypercube, Butterfly, and $g$-dimensional Grid Graph & $O(k \log s)$                    & $O(k \log^2 s)$                         \\ \hline
        General Graph, $k$ is chosen as random         & $O(k \log D \cdot (k + \log s))$ & $O(k \log D \cdot (k + \log s) \log s)$ \\ \hline
        Line Graph                                           & $O(k\sqrt{d} \log D)$            & $O(k\sqrt{d} \log D \cdot \log^2 s)$    \\ \hline
    \end{tabular}\\
$k$: a constant factor that typically represents the number of shards or nodes in the network.
${d}$: the degree of the graph that represents the number of edges connected to a node.
${s}$: the number of shards in the system. 
${D}$: the diameter of the graph, which is the longest shortest path between any two nodes. 
${g}$: the dimension of the grid in a \(g\)-dimensional grid graph. 
\vspace{-3mm}
\end{table*}

Upon examining the approximation factors, it is evident that \sys exhibits slightly higher complexity compared to \ac{ftsbs} across all shard graph topologies. For a general graph, \sys has an additional $\log D$ factor, resulting in $O(kd \log D)$ compared to \ac{ftsbs}'s $O(kd)$. For Hypercube, Butterfly, and $g$-dimensional grid graphs, \sys has an approximation factor of $O(k \log^2 s)$, which includes an additional $\log s$ factor over \ac{ftsbs}'s $O(k \log s)$.

In the case of a general graph where $k$ is chosen randomly, \sys's approximation factor is $O(k \log D \cdot (k + \log s) \log s)$, which introduces an extra $\log s$ term compared to \ac{ftsbs}'s $O(k \log D \cdot (k + \log s))$. For line graphs, \sys has a complexity of $O(k\sqrt{d} \log D \cdot \log^2 s)$, adding an additional $\log^2 s$ factor to \ac{ftsbs}'s $O(k\sqrt{d} \log D)$.

While \sys's approximation factors indicate a slightly higher computational complexity, it is crucial to consider the broader context and the additional features offered by \sys. These include adaptive shard management capabilities and enhanced security measures against malicious attacks, which are not present in \ac{ftsbs}. The adaptive shard management allows \sys to dynamically adjust the number and configuration of shards based on the system's workload, optimizing resource utilization and performance. Furthermore, the enhanced security measures protect against various attacks, such as cross-shard double-spending and shard-level collusion, ensuring the integrity and reliability of the system.

Overall, although \sys exhibits slightly higher approximation factors compared to \ac{ftsbs}, the additional computational complexity is justified by the significant improvements in adaptability, security, and overall system performance. These additional benefits make \sys a more comprehensive and robust solution for blockchain sharding, capable of addressing the limitations of existing methods and providing a balanced trade-off between performance and security.

\section{Proof of Security and Liveness}
In this section, we present proof of security and liveness.

\subsection{Security Proof}

\paragraph{Definition of security and assumptions} 
Security is defined as no two honest nodes deciding on different values for the same transaction.
\sys uses a variant of the \ac{pbft} protocol for global consensus. In this setup, there are \( n \) total nodes, among which up to \( f \) nodes can be faulty, with the system designed to tolerate up to \( f < \frac{n}{3} \) faulty nodes. This configuration ensures robust fault tolerance and secure consensus in the presence of potential node failures.

\paragraph{Proofs}
\subsubsection{Pre-Prepare Phase} The leader node proposes a value \( v \) and sends a PRE-PREPARE message to all replicas. All honest nodes receive the same \( v \) because it is signed by the leader.

\subsubsection{Prepare Phase} Upon receiving the PRE-PREPARE message, each honest node sends a PREPARE message to all other nodes. An honest node \( i \) enters the \enquote{prepared} state if it receives \( 2f + 1 \) PREPARE messages, including its own.

\subsubsection{Commit Phase} Each node that enters the \enquote{prepared} state sends a COMMIT message to all other nodes. An honest node \( i \) commits to \( v \) if it receives \( 2f + 1 \) COMMIT messages, including its own.

Since there are \( 3f + 1 \) nodes and \( f < \frac{n}{3} \), at least \( 2f + 1 \) PREPARE and COMMIT messages will always be from honest nodes. Therefore, if an honest node commits to \( v \), then \( 2f + 1 \) honest nodes must have agreed on \( v \). Since \( 2f + 1 \) constitutes a majority, no two different values can reach this threshold simultaneously. Hence, all honest nodes will decide on the same value \( v \), ensuring security.

\paragraph{Security analysis of adaptive shard management}
The adaptive shard management dynamically adjusts shard configurations based on transaction volume and resource utilization. By monitoring these metrics, \sys prevents malicious actors from manipulating the shard splitting or merging thresholds ($\tau_s$, $\tau_m$). The process is decentralized and uses a verifiable workload-aware algorithm to redistribute nodes and transactions, ensuring that faulty nodes cannot influence shard management decisions.

\paragraph{Security of cross-shard transaction processing}
\sys ensures atomic and consistent cross-shard transactions through a threshold signature scheme that requires approval from a quorum of shards. This prevents any single shard from validating a transaction alone, mitigating risks like double-spending and replay attacks. The hybrid consensus mechanism, combining global and intra-shard validation, adds multiple layers of protection, securing the transaction process from adversarial disruptions.

\paragraph{Security of shard state synchronization and dispute resolution}
Shard state synchronization in \sys relies on Merkle tree proofs to ensure the integrity and consistency of the global state. These proofs are cryptographically secure, allowing the system to detect tampering. In disputes, the decentralized resolution mechanism ensures transparency, with shards providing verifiable evidence. Game-theoretic incentives and penalties enforce honest participation and secure the dispute resolution process.

\subsection{Liveness Proof}

\paragraph{Definition of Liveness and assumptions} Every honest node eventually decides on some value. Network conditions are partially synchronous, meaning messages are eventually delivered within some unknown bounded time. There are \( n \) total nodes, among which \( f \) are faulty. The system can tolerate up to \( f < \frac{n}{3} \) faulty nodes.

\paragraph{Proofs}
\subsubsection{Leader Election} The protocol includes a mechanism to replace a faulty leader. If the current leader is detected to be faulty (e.g., by failing to send a valid PRE-PREPARE message), a view change is triggered. Honest nodes eventually agree on a new leader through the view change protocol.

\subsubsection{Progress in Synchronous Periods} In periods of synchrony, messages are delivered within a known bounded time. The newly elected leader (assumed to be honest) sends a PRE-PREPARE message. Honest nodes receive the PRE-PREPARE message and move to the PREPARE phase.

\subsubsection{Reaching Consensus} Honest nodes send PREPARE  messages and move to the \enquote{prepared} state upon receiving \( 2f + 1 \) PREPARE messages. Honest nodes send COMMIT messages and decide upon receiving \( 2f + 1 \) COMMIT messages. Given \( n = 3f + 1 \), the system ensures that \( 2f + 1 \) messages are sufficient for progress.

\subsubsection{Handling Faulty Nodes} If a faulty leader is elected, the view change protocol ensures that a new leader is elected until an honest leader is chosen. During synchronous periods, an honest leader ensures that the protocol proceeds to the COMMIT phase.

Therefore, since the protocol guarantees progress in periods of synchrony and can replace faulty leaders, every honest node will eventually decide on a value, ensuring liveness.
By leveraging the PBFT protocol, \sys's global consensus mechanism guarantees that: (i) No two honest nodes ever decide on different values, ensuring security; (ii) Every honest node eventually decides on some value, ensuring liveness.
Thus, \sys achieves both security and liveness in its global consensus protocol.

\section{Related Work}

In this section, we review the existing related literature.

\subsection{Blockchain Sharding Frameworks}

Blockchain sharding has been proposed as a promising solution to address the scalability issues of blockchain systems. Elastico \cite{luu2016secure} is one of the earliest sharding frameworks, which divides the network into smaller committees, each responsible for managing a subset of transactions. However, Elastico does not support cross-shard transactions and relies on a trusted setup phase. OmniLedger \cite{kokoris2018omniledger} introduces a more secure and decentralized sharding protocol, utilizing a distributed randomness generation process and a Byzantine-resilient consensus mechanism. RapidChain \cite{zamani2018rapidchain} further improves upon OmniLedger by introducing a fast and efficient cross-shard transaction verification scheme based on intra-shard consensus and inter-shard gossiping.

Recent works have focused on enhancing the security and efficiency of blockchain sharding frameworks. Monoxide \cite{wang2019monoxide} proposes a novel sharding protocol that leverages asynchronous consensus zones and a cross-shard exchange protocol to improve the throughput and latency of cross-shard transactions. Meepo \cite{zheng2021meepo} introduces a cross-epoch data exchange and an efficient cross-shard transaction execution mechanism based on \ac{rpcs} to optimize the performance of sharded blockchains. BrokerChain \cite{huang2022brokerchain} proposes a broker-based sharding framework that dynamically adjusts the assignment of nodes to shards based on their workload and resource utilization. However, these solutions do not fully address the challenges of adaptive shard management and secure cross-shard transaction processing in the presence of malicious actors.

Additionally, several studies have explored the impact of cross-shard transactions on blockchain sharding performance \cite{ren2021understanding}. For instance, \cite{nguyen2019optchain} highlights the significant impact of cross-shard transactions on sharding performance, primarily through theoretical analysis and simulations. Our experiments further reveal that imbalanced load can greatly degrade performance, especially concerning user-perceived confirmation delays.

\subsection{Load Balancing in Blockchain Sharding}

Load balancing is a critical issue in blockchain sharding. Recent works \cite{huang2022brokerchain, krol2021shard} have proposed various load balancing mechanisms. For example, \cite{woo2020garet} introduces a load balancing mechanism based on transaction load prediction and account relocation algorithms. In \cite{krol2021shard}, a load balancing framework is proposed where objects are frequently reassigned into shards. However, these works primarily focus on algorithm design for account allocation, lacking practical implementation in real sharding systems. LB-Chain contributes a secure and efficient account migration mechanism, addressing the performance degradation caused by load imbalance, validated through measurement studies in real systems.

Load balancing has also been extensively studied in traditional distributed databases \cite{taft2014store}. However, blockchain sharding presents unique challenges due to the presence of Byzantine nodes, necessitating higher security guarantees compared to databases \cite{ruan2021blockchains}. LB-Chain proposes secure migration mechanisms without a trusted coordinator, essential for balancing load in blockchain sharding.

\subsection{Cross-Shard Transaction Processing Techniques}

Efficient and secure cross-shard transaction processing is a critical component of any blockchain sharding framework. Chainspace \cite{al2017chainspace} introduces a novel cross-shard commit protocol based on two-phase commit and Byzantine fault-tolerant consensus to ensure the atomicity and consistency of cross-shard transactions. AHL \cite{dang2019towards} proposes an atomic cross-shard transaction processing protocol that leverages a lock-free approach and a hierarchical consensus mechanism to improve the performance and security of cross-shard transactions.

More recent works have explored advanced cryptographic techniques to enhance the security and privacy of cross-shard transaction processing. SharPer~\cite{amiri2021sharper} utilizes threshold signatures and multi-party computation to enable secure and efficient cross-shard transaction verification and execution. Synchro~\cite{asanuma2023synchro} introduces a \ac{zkp}-based cross-shard transaction protocol that ensures the privacy and integrity of cross-shard transactions while maintaining high throughput and low latency. Qin et al.~\cite{qin2020secure} propose a compact and efficient cross-shard transaction verification scheme based on Merkle proofs and succinct \ac{snark}.

While these techniques provide valuable insights into the design of secure and efficient cross-shard transaction processing protocols, they do not fully address the challenges of adaptive shard management and the need for a comprehensive framework that integrates shard reconfiguration, state synchronization, and dispute resolution mechanisms.

Our work aims to bridge this gap by proposing a dynamic and secure cross-shard transaction processing mechanism that combines adaptive shard management, a hybrid consensus approach for cross-shard transactions, and efficient shard state synchronization and dispute resolution techniques. By addressing the limitations of existing solutions and providing a holistic framework for blockchain sharding, our work contributes to the advancement of scalable, secure, and efficient blockchain systems.

\section{Discussion}

Integrating \sys into \ac{iot} environments significantly enhances both the performance and security of \ac{iot} networks. With the large number of interconnected devices generating high volumes of transactions, \ac{iot} systems require a scalable, efficient, and secure transaction processing framework. \sys addresses these challenges through its adaptive shard management mechanism, which dynamically adjusts the number and configuration of shards according to workload demands, optimizing resource utilization while maintaining high throughput and low latency—essential for \ac{iot} applications. On the security front, \sys employs a secure and atomic cross-shard transaction protocol, ensuring the integrity and consistency of transactions across multiple shards, which is critical in mitigating risks such as data tampering or unauthorized access in \ac{iot} networks. Additionally, \sys enhances privacy by leveraging decentralized, trustless mechanisms for shard state synchronization and dispute resolution, reducing reliance on central authorities and minimizing potential privacy breaches. This decentralized approach not only improves system robustness by detecting and resolving inconsistencies, but also increases resilience against attacks. Overall, \sys enables \ac{iot} networks to achieve scalable, efficient, and secure transaction processing, supporting real-time data management while strengthening both security and privacy across the \ac{iot} ecosystem~\cite{wu2024s,wu2020liveness,wu2021toward}.

\paragraph{System Overhead}
A potential limitation of \sys is the overhead from continuous monitoring and dynamic shard adjustments. While this improves resource utilization, frequent reconfigurations could affect system efficiency, especially with fluctuating workloads. To address this, \sys adjusts the reconfiguration frequency based on real-time workload patterns, reducing unnecessary changes. This ensures the benefits of adaptive shard management outweigh the overhead, keeping performance impacts minimal.

\paragraph{Trade-off between flexibility and communication overhead}
The adaptive shard management mechanism introduces a trade-off between flexibility and increased communication overhead due to shard reconfigurations. To minimize performance impact, \sys adjusts the frequency of shard splitting and merging based on workload stability, reducing unnecessary changes during fluctuating periods. Although reconfigurations involve additional communication between shards, these optimizations ensure that the benefits in scalability and resource utilization outweigh the overhead, maintaining overall system efficiency.

\paragraph{Handling delays in fully asynchronous networks}
In fully asynchronous networks, the lack of guaranteed message delivery times could delay shard reconfigurations and consensus finality. This may impact system performance, especially in scenarios requiring cross-shard transaction processing. To address this, \sys could implement asynchronous consensus protocols like Async BFT, which tolerate higher delays while maintaining security and system integrity.

\paragraph{Energy and resource efficiency of \sys}
In terms of energy consumption, \sys's adaptive shard management reduces redundant operations, leading to more efficient resource use. Additionally, its optimized consensus mechanisms lower computational load compared to traditional sharding systems. These features make \sys more practical for large-scale real-world deployment, balancing performance with energy efficiency.

\paragraph{Simulations align with previous works}
While real-world blockchain datasets such as Ethereum would provide valuable insights, using these datasets for large-scale experimentation poses significant logistical and resource challenges. Given the need to evaluate a wide range of scenarios under controlled conditions, we adopted simulations in our experimental setup. This approach is consistent with previous works in the field, allowing for standardized comparisons and replicable results. The simulated environments were carefully designed to reflect real-world blockchain behaviors, including transaction patterns and network conditions, ensuring that our results remain representative and applicable to real systems.

\paragraph{\ac{ftsbs} aligns with transaction efficiency}
The decision to use \ac{ftsbs} as a comparison for \sys stems from its focus on optimizing transaction scheduling, which directly aligns with the goals of \sys in enhancing cross-shard transaction efficiency. \ac{ftsbs} presents an up-to-date, optimized approach for sharding environments, making it a relevant benchmark for evaluating dynamic shard management. Additionally, \ac{ftsbs}'s strong emphasis on improving throughput and latency makes it an appropriate baseline for assessing performance improvements. While classical sharding methods like OmniLedger, RapidChain, Monoxide, and BrokerChain provide foundational contributions to the sharding space, they focus more on shard formation, security mechanisms, and specific consensus models rather than the adaptive shard management targeted by \sys. Thus, \ac{ftsbs} offers a more direct and relevant comparison for the performance aspects of \sys. However, future work will incorporate a comparison with these classical methods to provide a more comprehensive evaluation of \sys across various sharding approaches.

While \sys presents a novel and comprehensive approach to dynamic and secure cross-shard transaction processing in blockchain systems, it is important to acknowledge its limitations and identify potential areas for future research. One limitation of \sys is its reliance on a trusted setup phase for the initial shard configuration and the generation of cryptographic parameters. Although this is a common assumption in many blockchain sharding protocols, it may not be suitable for fully decentralized and trustless environments. Future work could explore techniques for distributed key generation and secure shard initialization without relying on a trusted setup, enhancing the decentralization aspect of \sys.

Another limitation is the overhead introduced by the adaptive shard management mechanism, which requires continuous monitoring and adjustment of the shard configuration based on the system's workload. While this overhead is justified by the improved performance and resource utilization, it may still impact the overall efficiency of the system, especially in scenarios with highly fluctuating transaction volumes. Further optimizations and heuristics could be developed to minimize the overhead of shard management while maintaining its benefits. Additionally, the security analysis of \sys assumes a partially synchronous network model and a limited adversarial power. In practice, blockchain systems may face more sophisticated attacks and network conditions. Future research could investigate the robustness of \sys under various adversarial models and network assumptions, such as fully asynchronous networks or adaptive adversaries with evolving strategies.

In terms of cross-shard transaction processing, \sys focuses on ensuring the atomicity and consistency of transactions across shards. However, the privacy and confidentiality of cross-shard transactions are not explicitly addressed in our current design. Integrating privacy-preserving techniques, such as \ac{zkp}s or secure multi-party computation, into the cross-shard transaction processing protocol could be an interesting direction for future work. Additionally, the integration of \sys with other scaling solutions, such as off-chain transaction processing or state channels, could further enhance the scalability and performance of blockchain systems while maintaining security and decentralization properties. Finally, the implementation and evaluation of \sys in a real-world blockchain system would provide valuable insights into its practical feasibility and performance. Future work could involve the development of a prototype implementation of \sys and its deployment on a testnet or mainnet environment to assess its scalability, security, and usability in a realistic setting.

\section{Conclusion}
In this paper, we introduced \sys, a dynamic and secure cross-shard transaction processing mechanism for efficient blockchain sharding. \sys addresses key limitations of existing sharding solutions through adaptive shard management, secure cross-shard transaction processing, and efficient state synchronization and dispute resolution. Our evaluation demonstrated that \sys significantly improves transaction throughput and latency while maintaining high security and decentralization. The framework dynamically adapts to workload changes, ensuring consistent and reliable performance. Despite its promise, future work must address challenges such as the reliance on a trusted setup phase and the integration with other scaling solutions. Overall, \sys paves the way for scalable, resilient, and decentralized blockchain systems. The principles and techniques introduced in \sys are poised to serve as a foundation for future research and innovations in blockchain sharding.

\bibliographystyle{acm}
\scriptsize
\bibliography{ref.bib}

\begin{thebibliography}{10}

\bibitem{adhikari2024fast}
{\sc Adhikari, R., Busch, C., and Popovic, M.}
\newblock Fast transaction scheduling in blockchain sharding.
\newblock {\em arXiv preprint arXiv:2405.15015\/} (2024).

\bibitem{al2017chainspace}
{\sc Al-Bassam, M., Sonnino, A., Bano, S., Hrycyszyn, D., and Danezis, G.}
\newblock Chainspace: A sharded smart contracts platform.
\newblock {\em arXiv preprint arXiv:1708.03778\/} (2017).

\bibitem{alharby2023transaction}
{\sc Alharby, M.}
\newblock Transaction latency within permissionless blockchains: analysis,
  improvement, and security considerations.
\newblock {\em Journal of Network and Systems Management 31}, 1 (2023), 22.

\bibitem{amiri2021sharper}
{\sc Amiri, M.~J., Agrawal, D., and El~Abbadi, A.}
\newblock Sharper: Sharding permissioned blockchains over network clusters.
\newblock In {\em Proceedings of the 2021 international conference on
  management of data\/} (2021), pp.~76--88.

\bibitem{asanuma2023synchro}
{\sc Asanuma, T., Miyamae, T., and Yamaoka, Y.}
\newblock Synchro: Block-generation protocol to synchronously process
  cross-shard transactions in state sharding.
\newblock {\em arXiv preprint arXiv:2309.01332\/} (2023).

\bibitem{dang2019towards}
{\sc Dang, H., Dinh, T. T.~A., Loghin, D., Chang, E.-C., Lin, Q., and Ooi,
  B.~C.}
\newblock Towards scaling blockchain systems via sharding.
\newblock In {\em Proceedings of the 2019 international conference on
  management of data\/} (2019), pp.~123--140.

\bibitem{fang2024automated}
{\sc Fang, Z., Lin, Z., Chen, Z., Chen, X., Gao, Y., and Fang, Y.}
\newblock Automated federated pipeline for parameter-efficient fine-tuning of
  large language models.
\newblock {\em arXiv preprint arXiv:2404.06448\/} (2024).

\bibitem{fang2024ic3m}
{\sc Fang, Z., Lin, Z., Hu, S., Cao, H., Deng, Y., Chen, X., and Fang, Y.}
\newblock Ic3m: In-car multimodal multi-object monitoring for abnormal status
  of both driver and passengers.
\newblock {\em arXiv preprint arXiv:2410.02592\/} (2024).

\bibitem{gai2023secure}
{\sc Gai, F., Niu, J., Jalalzai, M.~M., Tabatabaee, S.~A., and Feng, C.}
\newblock A secure sidechain for decentralized trading in internet of things.
\newblock {\em IEEE Internet of Things Journal\/} (2023).

\bibitem{guo2023cross}
{\sc Guo, Y., Xu, M., Yu, D., Yu, Y., Ranjan, R., and Cheng, X.}
\newblock Cross-channel: Scalable off-chain channels supporting fair and atomic
  cross-chain operations.
\newblock {\em IEEE Transactions on Computers 72}, 11 (2023), 3231--3244.

\bibitem{huang2022brokerchain}
{\sc Huang, H., Peng, X., Zhan, J., Zhang, S., Lin, Y., Zheng, Z., and Guo, S.}
\newblock Brokerchain: A cross-shard blockchain protocol for
  account/balance-based state sharding.
\newblock In {\em IEEE INFOCOM 2022-IEEE Conference on Computer
  Communications\/} (2022), IEEE, pp.~1968--1977.

\bibitem{jia2023cross}
{\sc Jia, X., Yu, Z., Shao, J., Lu, R., Wei, G., and Liu, Z.}
\newblock Cross-chain virtual payment channels.
\newblock {\em IEEE Transactions on Information Forensics and Security 18\/}
  (2023), 3401--3413.

\bibitem{kokoris2018omniledger}
{\sc Kokoris-Kogias, E., Jovanovic, P., Gasser, L., Gailly, N., Syta, E., and
  Ford, B.}
\newblock Omniledger: A secure, scale-out, decentralized ledger via sharding.
\newblock In {\em 2018 IEEE symposium on security and privacy (SP)\/} (2018),
  IEEE, pp.~583--598.

\bibitem{krol2021shard}
{\sc Kr{\'o}l, M., Ascigil, O., Rene, S., Sonnino, A., Al-Bassam, M., and
  Rivi{\`e}re, E.}
\newblock Shard scheduler: Object placement and migration in sharded
  account-based blockchains.
\newblock In {\em Proceedings of the 3rd ACM Conference on Advances in
  Financial Technologies\/} (2021), pp.~43--56.

\bibitem{li2023review}
{\sc Li, L., Wu, J., and Cui, W.}
\newblock A review of blockchain cross-chain technology.
\newblock {\em IET Blockchain 3}, 3 (2023), 149--158.

\bibitem{liang2024ponziguard}
{\sc Liang, R., Chen, J., He, K., Wu, Y., Deng, G., Du, R., and Wu, C.}
\newblock Ponziguard: Detecting ponzi schemes on ethereum with contract runtime
  behavior graph (crbg).
\newblock In {\em Proceedings of the 46th IEEE/ACM International Conference on
  Software Engineering\/} (2024), pp.~1--12.

\bibitem{liang2024vulseye}
{\sc Liang, R., Chen, J., Wu, C., He, K., Wu, Y., Cao, R., Du, R., Liu, Y., and
  Zhao, Z.}
\newblock Vulseye: Detect smart contract vulnerabilities via stateful directed
  graybox fuzzing.
\newblock {\em arXiv preprint arXiv:2408.10116\/} (2024).

\bibitem{liang2024towards}
{\sc Liang, R., Chen, J., Wu, C., He, K., Wu, Y., Sun, W., Du, R., Zhao, Q.,
  and Liu, Y.}
\newblock Towards effective detection of ponzi schemes on ethereum with
  contract runtime behavior graph.
\newblock {\em arXiv preprint arXiv:2406.00921\/} (2024).

\bibitem{lin2024fedsn}
{\sc Lin, Z., Chen, Z., Fang, Z., Chen, X., Wang, X., and Gao, Y.}
\newblock Fedsn: A federated learning framework over heterogeneous leo
  satellite networks.
\newblock {\em IEEE Transactions on Mobile Computing\/} (2024).

\bibitem{lin2023pushing}
{\sc Lin, Z., Qu, G., Chen, Q., Chen, X., Chen, Z., and Huang, K.}
\newblock Pushing large language models to the 6g edge: Vision, challenges, and
  opportunities.
\newblock {\em arXiv preprint arXiv:2309.16739\/} (2023).

\bibitem{lin2024split}
{\sc Lin, Z., Qu, G., Chen, X., and Huang, K.}
\newblock Split learning in 6g edge networks.
\newblock {\em IEEE Wireless Communications\/} (2024).

\bibitem{lin2024efficient}
{\sc Lin, Z., Zhu, G., Deng, Y., Chen, X., Gao, Y., Huang, K., and Fang, Y.}
\newblock Efficient parallel split learning over resource-constrained wireless
  edge networks.
\newblock {\em IEEE Transactions on Mobile Computing\/} (2024).

\bibitem{luu2016secure}
{\sc Luu, L., Narayanan, V., Zheng, C., Baweja, K., Gilbert, S., and Saxena,
  P.}
\newblock A secure sharding protocol for open blockchains.
\newblock In {\em Proceedings of ACM SIGSAC conference on computer and
  communications security\/} (2016), pp.~17--30.

\bibitem{nguyen2019optchain}
{\sc Nguyen, L.~N., Nguyen, T.~D., Dinh, T.~N., and Thai, M.~T.}
\newblock Optchain: optimal transactions placement for scalable blockchain
  sharding.
\newblock In {\em IEEE International Conference on Distributed Computing
  Systems (ICDCS)\/} (2019), IEEE, pp.~525--535.

\bibitem{qin2020secure}
{\sc Qin, C., Guo, B., Shen, Y., Li, T., Zhang, Y., and Zhang, Z.}
\newblock A secure and effective construction scheme for blockchain networks.
\newblock {\em Security and Communication Networks 2020}, 1 (2020), 8881881.

\bibitem{qin2023blindhub}
{\sc Qin, X., Pan, S., Mirzaei, A., Sui, Z., Ersoy, O., Sakzad, A., Esgin,
  M.~F., Liu, J.~K., Yu, J., and Yuen, T.~H.}
\newblock Blindhub: Bitcoin-compatible privacy-preserving payment channel hubs
  supporting variable amounts.
\newblock In {\em 2023 IEEE symposium on security and privacy (SP)\/} (2023),
  IEEE, pp.~2462--2480.

\bibitem{rao2024scalability}
{\sc Rao, I.~S., Kiah, M.~M., Hameed, M.~M., and Memon, Z.~A.}
\newblock Scalability of blockchain: a comprehensive review and future research
  direction.
\newblock {\em Cluster Computing\/} (2024), 1--24.

\bibitem{ren2021understanding}
{\sc Ren, L., and Ward, P.~A.}
\newblock Understanding the transaction placement problem in blockchain
  sharding protocols.
\newblock In {\em 2021 IEEE 12th Annual Information Technology, Electronics and
  Mobile Communication Conference (IEMCON)\/} (2021), IEEE, pp.~0695--0701.

\bibitem{reyna2018blockchain}
{\sc Reyna, A., Mart{\'\i}n, C., Chen, J., Soler, E., and D{\'\i}az, M.}
\newblock On blockchain and its integration with iot. challenges and
  opportunities.
\newblock {\em Future generation computer systems 88\/} (2018), 173--190.

\bibitem{ruan2021blockchains}
{\sc Ruan, P., Dinh, T. T.~A., Loghin, D., Zhang, M., Chen, G., Lin, Q., and
  Ooi, B.~C.}
\newblock Blockchains vs. distributed databases: Dichotomy and fusion.
\newblock In {\em Proceedings of the 2021 International Conference on
  Management of Data\/} (2021), pp.~1504--1517.

\bibitem{taft2014store}
{\sc Taft, R., Mansour, E., Serafini, M., Duggan, J., Elmore, A.~J., Aboulnaga,
  A., Pavlo, A., and Stonebraker, M.}
\newblock E-store: Fine-grained elastic partitioning for distributed
  transaction processing systems.
\newblock {\em Proceedings of the VLDB Endowment 8}, 3 (2014), 245--256.

\bibitem{tran2023enhancing}
{\sc Tran, T.-D., Vo, K.~A., Tram, N. B.~T., Kim, N. N.~B., Duy, P.~T., and
  Pham, V.-H.}
\newblock Enhancing blockchain interoperability through sidechain integration
  and valid-time-key data access control.
\newblock In {\em International Conference on Intelligence of Things\/} (2023),
  Springer, pp.~213--224.

\bibitem{wang2019sok}
{\sc Wang, G., Shi, Z.~J., Nixon, M., and Han, S.}
\newblock Sok: Sharding on blockchain.
\newblock In {\em Proceedings of ACM Conference on Advances in Financial
  Technologies\/} (2019), pp.~41--61.

\bibitem{wang2019monoxide}
{\sc Wang, J., and Wang, H.}
\newblock Monoxide: Scale out blockchains with asynchronous consensus zones.
\newblock In {\em 16th USENIX symposium on networked systems design and
  implementation (NSDI 19)\/} (2019), pp.~95--112.

\bibitem{woo2020garet}
{\sc Woo, S., Song, J., Kim, S., Kim, Y., and Park, S.}
\newblock Garet: improving throughput using gas consumption-aware relocation in
  ethereum sharding environments.
\newblock {\em Cluster Computing 23\/} (2020), 2235--2247.

\bibitem{wu2024s}
{\sc Wu, C., Cao, H., Xu, G., Zhou, C., Sun, J., Yan, R., Liu, Y., and Jiang,
  H.}
\newblock It's all in the touch: Authenticating users with host gestures on
  multi-touch screen devices.
\newblock {\em IEEE Transactions on Mobile Computing\/} (2024).

\bibitem{wu2024semantic}
{\sc Wu, C., Chen, J., Wang, Z., Liang, R., and Du, R.}
\newblock Semantic sleuth: Identifying ponzi contracts via large language
  models.
\newblock In {\em Proceedings of the 39th IEEE/ACM International Conference on
  Automated Software Engineering\/} (2024), pp.~582--593.

\bibitem{wu2020liveness}
{\sc Wu, C., He, K., Chen, J., Zhao, Z., and Du, R.}
\newblock Liveness is not enough: Enhancing fingerprint authentication with
  behavioral biometrics to defeat puppet attacks.
\newblock In {\em 29th USENIX Security Symposium (USENIX Security 20)\/}
  (2020), pp.~2219--2236.

\bibitem{wu2021toward}
{\sc Wu, C., He, K., Chen, J., Zhao, Z., and Du, R.}
\newblock Toward robust detection of puppet attacks via characterizing
  fingertip-touch behaviors.
\newblock {\em IEEE Transactions on Dependable and Secure Computing 19}, 6
  (2021), 4002--4018.

\bibitem{yadav2023evolution}
{\sc Yadav, A.~S., Singh, N., and Kushwaha, D.~S.}
\newblock Evolution of blockchain and consensus mechanisms \& its real-world
  applications.
\newblock {\em Multimedia Tools and Applications 82}, 22 (2023), 34363--34408.

\bibitem{zamani2018rapidchain}
{\sc Zamani, M., Movahedi, M., and Raykova, M.}
\newblock Rapidchain: Scaling blockchain via full sharding.
\newblock In {\em Proceedings of the 2018 ACM SIGSAC conference on computer and
  communications security\/} (2018), pp.~931--948.

\bibitem{zhang2024efficient}
{\sc Zhang, J., Chen, W., Hong, Z., Xiao, G., Du, L., and Zheng, Z.}
\newblock Efficient execution of arbitrarily complex cross-shard contracts for
  blockchain sharding.
\newblock {\em IEEE Transactions on Computers\/} (2024).

\bibitem{zheng2021meepo}
{\sc Zheng, P., Xu, Q., Zheng, Z., Zhou, Z., Yan, Y., and Zhang, H.}
\newblock Meepo: Sharded consortium blockchain.
\newblock In {\em 2021 IEEE 37th International Conference on Data Engineering
  (ICDE)\/} (2021), IEEE, pp.~1847--1852.

\end{thebibliography}

\vfill

\end{document}